\documentclass[journal]{IEEEtran}
\usepackage{amsfonts}
\usepackage{amssymb}
\usepackage[cmex10]{amsmath}
\usepackage{graphicx}
\usepackage{subfigure}
\usepackage{verbatim}
\usepackage{cite}
\usepackage{booktabs}
\usepackage[ruled]{algorithm2e}
\usepackage{multirow}
\usepackage{diagbox}
\usepackage{bm}
\usepackage{amsthm}

\newcommand{\bX}{{\mathbf X}}
\newcommand{\bx}{{\mathbf x}}

\newcommand{\bW}{{\mathbf W}}

\newcommand{\bPsi}{{\boldsymbol\Psi}}
\newcommand{\bpsi}{{\boldsymbol\psi}}
\newcommand{\bpi}{{\boldsymbol\pi}}

\newtheorem{Def}{Definition}
\newtheorem{Pro}{Proposition}

\newtheorem{Lem}{Lemma}

\allowdisplaybreaks

\ifCLASSINFOpdf
\else
\fi
\IEEEoverridecommandlockouts
\begin{document}
\title{Enhanced Approximation of Labeled Multi-object Density based on Correlation Analysis}

\author{
Wei Yi and Suqi Li\\
\IEEEauthorblockA{{University of Electronic Science and Technology of China, School of Electronic Engineering, Chengdu City, China} \\
{Email: kussoyi@gmail.com}}}


\maketitle

\IEEEpeerreviewmaketitle

\begin{abstract}
The multi-object density is a fundamental descriptor of a point process and has ability to describe  the randomness of the  number
 and the values of objects, as well as the statistical correlation  between objects. Due to its comprehensive nature, multi-object density usually has a complicate mathematical structure making
the set integral suffer from  the curse of dimension and the combinatorial nature of the problem. Hence, efficient and accurate  approximations of multi-object density is a key research theme in point process theory or finite
  set statistics. Conventional approaches usually discard  all or part of statistical correlation between objects mechanically in return
 for computational efficiency, without regard for the actual  correlation.
In this paper, we propose an enhanced approximation of the labeled multi-object (LMO)
 density by  adaptively factorizing the LMO density into  densities of several independent subsets based on the perception of the actual statistical correlation between object states. Besides, as a key process of obtaining a tractable factorization of LMO density, the labeled set marginal density of any subset suitable is derived for the universal labeled RFS, such as the generalized labeled multi-Bernoulli  RFS family and its subclasses.
The numerical studies show that the proposed approximation approach can significantly simplify the LMO density by utilizing the existing independence property while accurately reserve the statistical correlation between objects.

\end{abstract}
\section{Introduction}
In multi-object inference, the mission is to simultaneously estimate  the number of objects as well as their individual states. The applications of multi-object inference  spin over a wide range of areas, such as forestry \cite{refr:forest}, biology \cite{refr:biology}, physics \cite{refr:physics}, computer vision \cite{refr:computer-vision}, wireless
networks \cite{refr:network}, communications \cite{refr:communication}, multi-target
tracking \cite{refr:tracking-1,refr:tracking-2}, and robotic\cite{refr:robotics}. The
states of objects in multi-object systems, for instance the coordinates of molecules in a liquid, trees in a forest and stars in a galaxy, is a typical point pattern  modeled by point processes (specifically simple finite point processes or random finite sets (RFS)) derived from stochastic geometry. The point process theory \cite{refr:point-process} provides the tools for characterizing the underlying laws of the point patterns. Finite set statistics (FISST) \cite{refr:Mahler_book} proposed by Mahler also  provides
 mathematical tools for dealing with
RFSs based on a notion of integration and density that
is consistent with point process theory.

A fundamental descriptor of point processes is multi-object probability density which  captures the uncertainty of  the number and
values of objects, as well as the statistical correlation between objects. Due to its comprehensive nature, the multi-object density usually has a complicate mathematical structure, more specifically, the multiple hypotheses involving
 different cardinalities, and the high-dimensional densities conditional on given cardinalities.
The core of  multi-object estimation is dynamic Bayesian inference.  Computation of  the posterior density via Bayes rule requires the integration of the product of the prior density and likelihood function. This integration poses  practical challenges especially for multi-object probability prior because the complicate structure of multi-object probability
 density makes the set integral suffer from the curse of dimensionality and the inherently combinatorial nature of the problem.

To solve these problems, tractable approximations of multi-object probability density are necessary
and two points during the approximation should be remarked. Firstly, statistical independence between objects can be utilized to enable
 the parallel implementation to reduce both the number of combinations and  the dimension of joint density. Secondly, statistical correlation  between
 objects also should be reserved when dependence actually exists. Statistical correlation  usually comes from the ambiguous observation (relative to multiple objects)
 when considering the posterior multi-object density, or from the interactions between objects in Markov point processes \cite{refr:Markov_PP}. For instance, when  the objects are in proximity, there remain large uncertainty for the association map between object states and its observations for standard observation model, or the object superpositions arise  for image observation.  Then multi-object posterior should admit statistical correlation between objects. Ignoring the statistical correlation is likely to lead to an estimation statistical bias during the multi-object inference.

Conventional approaches usually approximate multi-object probability density as a certain class of density. There exist two categories of approximate densities: one is to  completely  discard the correlation between objects and assume thorough  independence of objects, such as Poisson process \cite{refr:Mahler_book,refr:PHD}, independent identically distributed (i.i.d.) process \cite{refr:Mahler_book,refr:CPHD},
 multi-Bernoulli (MB) density \cite{refr:Mahler_book,refr:MeMber_filter1,refr:MeMber_filter}. While this kind of densities enjoys many  analytical
properties, it has been shown that sometimes they are too simplistic for the dynamic
Bayesian inference of point processes in complicated scenarios \cite{refr:MeMber_filter}. The other is to  cast away only a part of correlation between objects with the typical examples  generalized labeled multi-Bernoulli (GLMB) \footnote{GLMB distribution is also simply named as Vo-Vo distribution by Malher in his book \cite{refr:tracking-2} first time. } RFS family and its subclasses \cite{refr:label_1,refr:label_2,refr:label_3} . The advantages of the class of GLMB density is that it is  a conjugate prior that is also closed under the Chapman-Kolmogorov equation for the standard multi-object system model. Moreover, the set integral of GLMB density only involves the integrals on single-object space thus getting rid of  the curse of dimensionality.  However, the class of GLMB densities are  not necessarily closed under generic multi-object system  \cite{refr:label_6} because it still assumes independence of objects under each hypothesis involving the existence of different objects. To solve this problem, $\delta$-GLMB density approximation of labeled multi-object (LMO) density is proposed to enable $\delta$-GLMB filter for generic multi-object system \cite{refr:label_6}.
To summarize, the conventional approximate approaches  usually discard  part or all of statistical correlation  mechanically in return
 for computational efficiency, without regard for the real situation of correlation between objects.


Recently, the labeled set filters have achieved great developments for both standard and  generic multi-object system. As a whole,
the advantages of labeled set filters compared to previous (unlabeled)
random set filters are that they can  produce target trajectories formally, and  simplify the standard multi-object transition kernel in terms of both notation and complexity. Motivated by the developments and advantages of labeled set filters, it is significant to explore the efficient and accurate approximation of LMO density.

In this paper, we propose an enhanced approximate approach for LMO density \cite{refr:label_6} which does approximation  based on the perception of the actual statistical properties between objects.
The proposed method does not follow the old routine to approximate the LMO density using a certain type of distribution
 mechanically regardless of the real correlation between objects. In contrast, it evaluates the correlation between objects adaptively  and factorizes the LMO density into densities of several independent subsets  according to correlation analysis.
 The proposed method takes into account the simplification of the complicate structure of  LMO density and the reservation of  correlation when necessary. The proposed method is designed for the universal LMO density,  so it is also applicable to the  small classes of LMO density, such as GLMB RFS family and its subclasses.


The key point of the proposed approximation is the labeled  set marginal density. However, the computation of  labeled set marginal density is not mature. In \cite{refr:JMB}, we preliminarily give the concept of set marginal and its computing method for joint multi-Bernoulli RFS.  In this paper, we further derive the analytical expressions of set marginal density for the
  universal LMO density, GLMB density family and some subclasses of GLMB density including $\delta$-GLMB and Marginal $\delta$-GLMB (M$\delta$-GLMB) density, which  guarantee the  proposed approximation has great practicability.

The paper is organized as follows, the background of this paper is presented in Section II. Section III proposes an enhanced approximation of LMO density which factorizes the LMO density based on the  perception of the actual statistical correlation between object states. Section IV demonstrates the proposed approximation approach via numerical examples. Conclusion remarks are given in Section V.

\section{Background}
\subsection{Notations}
We adhere to the convention that single-object states are
represented by lowercase letters, e.g., $\bx$, while multi-object
states are represented by uppercase letters, e.g., $\bX$, $X$.  To distinguish labeled states and distributions from the
unlabeled ones, bold-type letters are adopted for the labeled
ones, e.g., $\bx$, $\bX$, $\bpi$. Moreover, blackboard bold letters represent spaces, e.g., the
state space is represented by $\mathbb{X}$, the label space by $\mathbb{L}$. The collection of all finite sets of $\mathbb{X}$
is denoted by $\mathcal{F}(\mathbb{X})$.

We use the multi-object exponential notation
\begin{equation}\label{multi-object exponential notation }
  h^{X}\triangleq\prod_{x\in X}h(x)
\end{equation}
for real-valued function $h$, with $h^\emptyset=1$ by convention.
To admit arbitrary arguments like sets, vectors and integers, the generalized Kronecker delta function is given by
\begin{equation}\label{delta}
  \delta_Y(X)\triangleq\left\{\begin{array}{l}
\!\!1,\,\,\, \mbox{if}\,\,\,X=Y\\
\!\!0,\,\,\, \mbox{otherwise}.
\end{array}\right.
\end{equation}
The inclusion function $1_Y(X)$ is given by
\begin{equation}\label{eq:A2}
  1_Y(X)=\left\{ \begin{array}{ll}
\ 1, & \mbox{if}\,\,X\subseteq Y\\
\ 0, & \mbox{otherwise}.\\
\end{array} \right.
\end{equation}
If $X$ is a singleton, i.e., $X=\{x\}$, the notation $1_Y(\{x\})$ is used instead of $1_Y{\{x\}}$.
\subsection{Labeled RFS and LMO Density}
A \textit{labeled RFS} is an RFS whose elements are identified by distinct labels \cite{refr:label_1,refr:label_2}. A \textit{labeled RFS} with (kinematic) state space $\mathbb{X}$ and (discrete) label space
 $\mathbb{L}$ is an RFS on $\mathbb{X}\times\mathbb{L}$ such that each realization $\mathbf{X}$ has distinct labels. Namely, a labeled RFS and the set of its labels have the same cardinality, $|\mathcal{L}(\mathbf{X})|=|\mathbf{X}|$, where $\mathcal{L}(\bX)=\{\mathcal{L}(\bx),\bx\in\bX\}$ denotes the set of labels of $\bX$ with $\mathcal{L}((x,\ell))=\ell$. A labeled RFS and its unlabeled
 version have the same cardinality distribution. For an arbitrary labeled RFS, its multi-object density can be represented as the expression given in Lemma 1
 \cite{ refr:Angel_1,refr:label_6}, and our main results in this paper follow from this
 expression.
 \begin{Lem} Given an LMO density $\bpi$ on $\mathcal{F}(\mathbb{X}\times\mathbb{L})$, and for any positive integer $n$, we define the joint existence probability of the label set $\{\ell_1,\ell_2,\cdots,\ell_n\}$ by
\begin{equation}\label{joint existence probability}
  \omega(\{\ell_1,\cdots,\ell_n\})\!=\!\int\!\! \bpi(\{(x_1,\ell_1),\!\cdots,\!(x_n,\ell_n)\})d(x_1,\!\cdots\!,x_n)
\end{equation}
and the joint probability density on $\mathbb{X}^n$ of the states $x_1,\cdots,x_n$ conditional on their corresponding labels $\ell_1,\cdots,\ell_n$ by
\begin{equation}\label{joint probability density}
  p(\{(x_1,\ell_1),\cdots,(x_n,\ell_n)\})=\frac{\bpi(\{(x_1,\ell_1),\cdots,(x_n,\ell_n)\})}{\omega(\{\ell_n,\cdots,\ell_n\})}
\end{equation}
Thus, the LMO density can be expressed as
\begin{equation}\label{factorized}
  \bpi(\mathbf{X})=\omega(\mathcal{L}(\mathbf{X}))p(\mathbf{X}).
\end{equation}
\end{Lem}



\subsection{GLMB RFS Family and Its Subclasses}
GLMB RFS family  \cite{refr:label_1} is a class of tractable labeled RFSs whose densities are conjugate with standard multi-object likelihood function, and are closed under the multi-object Chapman-Kolmogorov equation with respect
to the standard multi-object transition kernel. In order to facilitate the development of applications in signal processing
and related fields, \cite{refr:label_1}  introduces a smaller family within the
class of GLMB RFSs that is also
closed under the Chapman-Kolmogorov equation and Bayes
rule, namely $\delta$-GLMB RFS, and develops the $\delta$-GLMB filter. Nevertheless both GLMB and $\delta$-GLMB filters exhibit an exponential growth in the number of posterior components. Therefore on one hand, efficient implementation techniques of GLMB filter and  $\delta$-GLMB filter  are proposed in \cite{refr:label_1} and \cite{refr:label_efficient}.  On the other hand, two principled  approximations of  $\delta$-GLMB density  i.e., labeled multi-Bernoulli (LMB)  and marginalized $\delta$-GLMB (M$\delta$-GLMB) densities, and the corresponding filters, are proposed in \cite{refr:label_3} and \cite{refr:label_5}. Note that  both LMB and M$\delta$-GLMB densities are the subclasses of GLMB RFS family. Here, we present the definitions of GLMB RFS and its subclasses.

A GLMB RFS is a labeled RFS with state space $\mathbb{X}$ and (discrete) label space $\mathbb{L}$  distributed according to
\begin{align}\label{GLMB}
\begin{split}
\bpi_{\text{GLMB}}(\bX)=\Delta(\bX)\sum_{c\in\mathbb{C}}w^{(c)}(\mathcal{L}(\bX))[p^{(c)}]^\bX
\end{split}
\end{align}
where $\mathbb{C}$ is a discrete index set, $w^{(c)}(L)$ and $p^{(c)}$ satisfy
\begin{align}
\begin{split}
\sum_{L\subseteq\mathbb{L}}\sum_{c\in\mathbb{C}}w^{(c)}(L)&=1\\
\int p^{(c)}(x,\ell)dx&=1.
\end{split}
\end{align}
and $\Delta(\bX)=\delta_{|\bX|}(|\mathcal{L}(\bX)|)$ is the distinct-label indicator of $\bX$.

An $\delta$-GLMB RFS with state space $\mathbb{X}$ and discrete label space $\mathbb{L}$ is a special case of GLMB RFS with
\begin{equation}\label{delta-GLMB}
\begin{split}
\mathbb{C}=&\mathcal{F}(\mathbb{L})\times\Xi\\
\omega^{(c)}(L)=&\omega^{(I,\xi)}(L)=\omega^{(I,\xi)}\delta_I(L)\\
p^{(c)}=&p^{(I,\xi)}=p^{(\xi)}
\end{split}
\end{equation}
where $\Xi$ is a discrete space, i.e., it is distributed according to
\begin{equation}\label{delta-GLMB}
\bpi_{\delta\text{-GLMB}}(\bX)=\Delta(\bX)\sum_{(I,\xi)\in\mathcal{F}(\mathbb{L})\times\Xi}\omega^{(I,\xi)}\delta_{I}(\mathcal{L}(\bX)){[p^{(\xi)}]}^\bX.
\end{equation}

An M$\delta$-GLMB density $\bpi_{\text{M}\delta\text{-GLMB}}$ corresponding to the $\delta$-GLMB density $\bpi_{\delta\text{-GLMB}}$ in (\ref{delta-GLMB}) is a probability
density of the form
\begin{equation}\label{Mdelta-GLMB}
\bpi_{\text{M}\delta\text{-GLMB}}(\bX)=\Delta(\bX)\sum_{I\in\mathcal{F}(\mathbb{L})}\omega^{(I)}\delta_{I}(\mathcal{L}(\bX)){[p^{(I)}]}^\bX
\end{equation}
where
\begin{equation}
\begin{split}
\omega^{(I)}=&\sum_{\xi\in\Xi}\omega^{(I,\xi)}\\
p^{(I)}(\bx,\ell)=&1_I(\ell)\frac{1}{\omega^{(I)}}\sum_{\xi\in\Xi}\omega^{(I,\xi)}p^{(\xi)}(\bx,\ell).
\end{split}
\end{equation}

A LMB RFS\cite{refr:label_5}  with state space $\mathbb{X}$, label space $\mathbb{L}$ and (finite) parameter set $\{(r^{(\ell)},p^{(\ell)}(x)):\ell\in\mathbb{L}\}$, is distributed according to
\begin{align}\label{LMB}
\begin{split}
\bpi(\mathbf{X})=\Delta(\bX)w(\mathcal{L}(\bX))p^\bX
\end{split}
\end{align}
where
\begin{align}
\begin{split}
w(L)&=\prod_{i\in\mathbb{L}}(1-r^{i})\prod_{\ell\in L }\frac{1_{\mathbb{L}}(\ell)r^{\ell}}{1-r^{\ell}}\\
p(x,\ell)&=p^{(\ell)}(x).
\end{split}
\end{align}
\subsection{$\delta$-GLMB Density Approximation of LMO Density}
An  arbitrary LMO density can be approximated as a tractable $\delta$-GLMB density based on Lemma 2 \cite{refr:label_6}.  The $\delta$-GLMB density  approximation shown in (\ref{GLMB approximation}) abandons the statistical correlation between states under each hypotheses (involving different label set $I$), and hence get rid of the curse of dimensionality.  The reasonability and efficiency of the $\delta$-GLMB density approximation has been demonstrated by the relavant $\delta$-GLMB filter for generic observation model \cite{refr:label_6}.
However, the   $\delta$-GLMB density  approximation does not consider the actual correlation between objects, and may have considerable approximation error when  the correlation between objects is strong.
In addition, $\delta$-GLMB density  still suffers from  combination nature of problem, because the number of  hypotheses still increases exponentially with maximum object number.

\begin{Lem}
Given any LMO density $\bpi$ of form (\ref{factorized}), the $\delta$-GLMB density which preserves the cardinality distribution and probability hypothesis density (PHD) of $\bpi$, and minimizes the Kullback-Leibler divergence from $\bpi$, is given by
\begin{equation}\label{GLMB approximation}
  \hat{\bpi}_{\delta\text{-GLMB}}(\bX)=\Delta(\bX)\sum_{I\in\mathcal{F}(\mathbb{L})}\hat{\omega}^{(I)}\delta_{I}(\mathcal{L}(\bX)){[\hat{p}^{(I)}]}^\bX
\end{equation}
where
\begin{equation}\label{GLMB_where}
\begin{split}
\hat\omega^{(I)}&=\omega(I)\\
\hat{p}^{(I)}(x,\ell)&=1_I(\ell)p_{I-\{\ell\}}(x,\ell)\\
p_{\{\ell_1,\cdots,\ell_n\}}(x,\ell)&=\\
\int p(\{(x,\ell),&(x_1,\ell_1),\cdots,(x_n,\ell_n)\})d(x_1,\cdots,x_n).
\end{split}
\end{equation}
\end{Lem}
\subsection{Correlation Coefficient}
The correlation coefficient \cite{refr:probability_theory} is a measure that determines the degree to which two random variables are correlated.
The most commonly used measure is the Pearson's correlation coefficient, or simply called ``the correlation coefficient''. The correlation coefficient between two random variables $A$ and $B$ with $\mbox{cov}(A,B)$ the covariance of $A$ and $B$, and $\sigma_A$, $\sigma_B$ the standard deviations, is defined as:
\begin{equation}\label{correlation coefficient}
\rho_{A,B}=\frac{\mbox{cov}(A,B)}{\sigma_{A}\sigma_{B}}.
\end{equation}
Our results in this paper follow from this correlation coefficient, which is sensitive only to a linear relationship between two variables. This correlation coefficient  is only  applicable to evaluate the correlation between variables modeled by a random vector, but cannot provide a reasonable definition  to the RFS because both the elements and cardinality of an RFS are random.
\subsection{Kullback-Leibler divergence}
In probability theory and information theory, the Kullback-Leibler divergence (KLD) \cite{refr:KLD_1} is a measure of the difference between two probability distributions, and its extension to multi-object densities
$f(X)$ and $g(X)$ is given in \cite{refr:KLD_2} by
\begin{equation}\label{KLD}
D_{KL}(f;g)=\int f(X)\log\frac{f(X)}{g(X)}\delta X
\end{equation}
where the integral in (\ref{KLD}) is a set integral.
\section{Enhanced Approximate Strategy based  on Correlation Analysis}
The LMO density usually is approximated as a given type of multi-object density. In \cite{refr:label_6}, it proposed a tractable $\delta$-GLMB density approximation for an arbitrary LMO density, which matches the PHD and cardinality distribution of LMO density.
For standard multi-object system, GLMB density is a closed solution \cite{refr:label_1}, and it can be further approximated using LMB density \cite{refr:label_5} or M$\delta$-GLMB density \cite{refr:label_4}.
These approximations usually discard all or part of correlation of original LMO density mechanically for the sake of computation efficiency, without respect for the real situation of correlation between objects.

Actually, correlation of objects play an important role  in multi-object estimation. On one hand, when objects exhibit no correlation, the statistical independence can be utilized to enable parallel implementation, and thus simplify computation and enhance estimation performance \cite{refr:label_4,refr:label_7,closely-spaced-1,closely-spaced-2,closely-spaced-3,closely-spaced-4,closely-spaced-5,closely-spaced-6}. On the other hand, when objects are strongly correlated with each other, their statistics should be jointly considered, or it will produce poor estimation like the the aforementioned strategies.

In practice, the real situation of correlation between objects is usually complicate.   Empirical data suggests that in most  scenarios  not all objects  have correlation with each other, but only a small faction of objects  has correlation with the other small faction of objects and which object has correlation with which is  usually unknown and time-varying.


In this section, we present an enhanced approximate strategy for the approximation of LMO density  in which the correlation between different objects is estimated adaptively,
and the original LMO density is decomposed into densities of several independent subsets according to the correlation estimate. Besides, we derive the analytical expression of labeled set marginal density of any subsets of the universal LMO RFS,  GLMB RFS family and its subclasses.
\subsection{Correlation Estimate and Grouping}
Firstly, we introduce a concept of\textit{ basic component } of labeled RFS in  Definition 1, which is important in estimating the correlation between objects.
\begin{Def}
For an  arbitrary labeled RFS $\bPsi$ on space $\mathbb{X}\times\mathbb{L}$ ($\mathbb{L}$ is a finite label space), it can be seen as the union of $|\mathbb{L}|$
random subsets, i.e., $\bPsi=\biguplus_{\ell\in\mathbb{L}}\bpsi_\ell$, with each $\bpsi_\ell$  on space $\mathbb{X}\times\{\ell\}$.
We refer each random finite subset $\bpsi_\ell, \ell\in\mathbb{L}$ to as  a basic component of $\bPsi$.
\end{Def}
A basic component  $\bpsi_\ell$, namely, the random finite subset related to the objects with label $\ell$,  is a labeled Bernoulli RFS which is either the empty set
or the singleton set $\{\bx,\ell\}$.

A basic component  $\bpsi_\ell$ is the mathematical representation of the object   $\ell$, thus evaluating the correlation between different objects $  \ell$ and $\ell'$ amounts to evaluating the correlation between different basic components $\bpsi_\ell$ and $\bpsi_{\ell'}$.  Generally, the dimension of  the random vector is fixed,   thus its basic random variables related to the single object can only describe the randomness of object state and  cannot   describe   the uncertainty of object existence. By contrast, the basic components constituting an RFS can accommodate both the uncertainty of existence and the randomness of  object state. Hence,  to evaluate the correlation between different basic components, we should consider comprehensively from two aspects: 1)  the correlation of objects' existences; 2) the correlation
of  object states.
\subsubsection{Absolute Correlation Coefficient of  Existence, $\alpha_{\ell,\ell'}$ }
To describe its uncertainty of existence, we define a random variable $E_\ell$ for each basic component $\bpsi_\ell$ as
\begin{equation}\label{E_definition}
E_\ell=\left\{\begin{array}{ll}
0,  &\bpsi_\ell=\emptyset\\
1,  &\bpsi_\ell=\{(\bx,\ell)\}
\end{array}\right., \,\,\,\ell\in\mathbb{L}
\end{equation}
and the statistics of all $E_\ell$s, $\ell\in\mathbb{L}$ are distributed according to the joint probability distribution
\begin{equation}\label{distribution of existence}
\Pr\left(\left(\cap_{\ell\in I}\{E_\ell=1\}\right)\cap\left(\cap_{\ell'\in \mathbb{L}/I}\{E_{\ell'}=0\}\right)\right)=\omega(I),\,\,\, I\subseteq\mathbb{L}
\end{equation}
where ``$/$'' denotes the different set, and $\omega(I)$ is the joint
existence probability of the label set $I$ given in Lemma 1.

We define  the \textit{absolute correlation coefficient of existence} between $\bpsi_\ell$ and $\bpsi_{\ell'}$, $\ell\neq\ell'\in\mathbb{L}$ as
\begin{equation}\label{correlation coefficient of existence}
\alpha_{\ell,\ell'}=\left|\rho_{E_\ell,E_{\ell'}}\right|
\end{equation}
where $|\cdot|$ denotes the absolute value of $\cdot$, and $\rho_{E_\ell,E_{\ell'}}$ is the correlation coefficient between  $E_\ell$ and $E_{\ell'}$.

$\rho_{E_\ell,E_{\ell'}}$ can be computed from the joint existence distribution in (\ref{distribution of existence}) according to (\ref{correlation coefficient}). The specified formula of $\rho_{E_\ell,E_{\ell'}}$ is
\begin{small}
\begin{equation}\label{formula of rhoE}
\begin{split}
&\rho_{E_\ell,E_{\ell'}}=\\
&\frac{\epsilon_{\ell,\ell'}^{(1,1)}-(\epsilon_{\ell,\ell'}^{(1,0)}+\epsilon_{\ell,\ell'}^{(1,1)})(\epsilon_{\ell,\ell'}^{(0,1)}+\epsilon_{\ell,\ell'}^{(1,1)})}{\sqrt{(\epsilon_{\ell,\ell'}^{(1,0)}\!+\!\epsilon_{\ell,\ell'}^{(1,1)})(1\!-\!\epsilon_{\ell,\ell'}^{(1,0)}\!-\!\epsilon_{\ell,\ell'}^{(1,1)})(\epsilon_{\ell,\ell'}^{(0,1)}\!+\!\epsilon_{\ell,\ell'}^{(1,1)})(1\!-\!\epsilon_{\ell,\ell'}^{(0,1)}\!-\!\epsilon_{\ell,\ell'}^{(1,1)})}}
\end{split}
\end{equation}
\end{small}
where
\begin{equation}\label{Ma_probability_E}
\epsilon^{(i,j)}_{\ell,\ell'}=\sum_{I \in\mathcal{F}(\mathbb{L})}\omega(I)h^{(I)}(\ell,i)h^{(I)}(\ell',j),\,\,\,\,\,\,\,\, i, j=0,1
\end{equation}
with
\begin{equation}
h^{(I)}(\ell,i)=(1-1_I(\ell))\delta_0(i)+1_I(\ell)\delta_1(i)
\end{equation}
The proof of (\ref{formula of rhoE}) is given in Appendix A.
\subsubsection{Absolute Correlation Coefficient of  State, $\beta_{\ell,\ell'}$}
 The reasonability and efficiency of the $\delta$-GLMB density approximation has been demonstrated by the relavant $\delta$-GLMB filter for generic observation model \cite{refr:label_6}.
The estimation of the correlation between states of $\bpsi_{\ell}$ and $\bpsi_{\ell'}$ is on the condition that both $\bpsi_{\ell}$ and $\bpsi_{\ell'}$  exist.
 Under each hypothesis (involving the existing objects with  label set $I\in\mathcal{F}(\mathbb{L})$) where  $I$ includes $\ell$ and $\ell'$, we can compute a correlation coefficient between the labeled states $(\bx,\ell)$ and $(\bx',\ell')$,
 denoted as $\rho_{\ell,\ell'|I}$,
from the corresponding conditional joint probability density $p(\bX)(\mathcal{L}(\bX)=I)$ defined in Lemma 1, according to (\ref{correlation coefficient}).

For any two $\ell\neq\ell'\in\mathbb{L}$, we define the \textit{absolute correlation coefficient of state }between $\bpsi_\ell$ and $\bpsi_{\ell'}$
as
 \begin{equation}\label{correlation coefficient of state}
 \begin{split}
& \beta_{\ell,\ell'}=\frac{\sum_{I\in\mathcal{F}(\mathbb{L})}1_{I}(\{\ell,\ell'\})\omega(I)\left|\rho_{\ell,\ell'|I}\right|}{\sum_{I\in\mathcal{F}(\mathbb{L})}1_{I}(\{\ell,\ell'\})\omega(I)}.
\end{split}
 \end{equation}
Essentially, actually $\beta_{\ell,\ell'}$ is the weighted sum of absolute $\rho_{\ell,\ell'|I}$s over all hypotheses where the  label set $I$ includes $\ell$ and $\ell'$.

%

\begin{Def}
For an  arbitrary labeled RFS $\bPsi$ on space $\mathbb{X}\times\mathbb{L}$ ($\mathbb{L}$ is a finite label space), the absolute correlation coefficient between any two basic components
$\bpsi_{\ell_1}$ and $\bpsi_{\ell_2}$, $\ell_1\neq\ell_2\in\mathbb{L}$, is defined as
\begin{equation}\label{correlation estimate}
 \gamma_{\ell_1,\ell_2}=\omega_E \alpha_{\ell_1,\ell_2}+ \omega_S\beta_{\ell_1,\ell_2}\end{equation}
where $\omega_E+\omega_S=1$ with $\omega_E$, $\omega_S$ the weighting coefficients of $\alpha_{\ell_1,\ell_2}$, $\beta_{\ell_1,\ell_2}$ respectively, and $\alpha_{\ell_1,\ell_2}$, $\beta_{\ell_1,\ell_2}$ are  the absolute correlation coefficients of  existence and state defined in (\ref{correlation coefficient of existence}) and (\ref{correlation coefficient of state}), respectively.
\end{Def}

Note that the absolute correlation coefficient whose value goes form 0 to 1 is  an indicator to evaluate the correlation between basic components comprehensively.
 The value is  bigger, the correlation is stronger and vice versa. The value of $\omega_E$ or $\omega_S$ varies with different applications. If the correlation of state is emphasized, the value of $\omega_S$ is larger; otherwise, the value of $\omega_E$ is larger.

 After estimating the correlation between different  basic components, we can divide all $\bpsi_\ell$s, $\ell\in\mathbb{L}$ into several groups such that basic components within a group exhibit  correlation,
and basic components between different groups are statistically independent in the sense of (\ref{correlation estimate}). We represent each group as the union of the basic components within
the group, then $\bPsi$ can be divided into several independent random finite subsets, i.e., $\bPsi=\biguplus_{i=1}^N\bPsi_i$, where $\biguplus$ denotes the disjoint union.

\subsection{ Factorization of LMO density}
For an arbitrary RFS (unlabeled or labeled version),  the relationship between its density and the densities of its independent subsets are presented in Lemma 3, which is derived in \cite{refr:Mahler_book}.
\begin{Lem}
Let $\Psi=\Psi_1\cup\cdots\cup\Psi_N$ where $\Psi_1,\cdots,\Psi_N$ are statistically independent random finite subsets. The probability density of $\Psi$ is related to
the probability densities of $\Psi_1,\cdots,\Psi_n$ as follows:
\begin{equation}
\pi_\Psi(X)=\sum_{W_1\uplus\cdots\uplus W_n=X} \pi_{\Psi_1}(W_1)\cdots \pi_{\Psi_n}(W_n).
\end{equation}
\end{Lem}

For the labeled RFS, the conclusion given in Lemma 3 can be further specified, as shown in Proposition 1 whose proof is given in Appendix B.
\begin{Pro}
If a labeled RFS $\bPsi$ on space $\mathbb{X}\times\mathbb{L}$ can be divided into $N$ independent label random subsets
$\bPsi_i$ on space $\mathbb{X}\times\mathbb{L}_i, i=1,\cdots, N$, i.e., $ \bPsi=\bigcup_{i=1}^N  \bPsi_i$ with $\mathbb{L}=\mathbb{L}_1\uplus\cdots\uplus\mathbb{L}_N$,
 then the probability density of $\bPsi$  is related to
the probability densities of $\bPsi_1,\cdots,\bPsi_n$ as follows
\begin{equation}\label{SMD-approx}
\bpi_{{\bPsi}}(\bX)=\bpi_{{\bPsi}_1}(\bX\cap\mathbb{X}\times\mathbb{L}_1)\cdots \bpi_{{\bPsi}_N}(\bX\cap\mathbb{X}\times\mathbb{L}_N).
\end{equation}
\end{Pro}
According to Proposition 1, one can  find that the LMO density can be decomposed into several  LMO densities of independent random finite subsets, according to the correlation analysis  based
grouping of basic components. As a result, we obtain  an approximation of LMO density which   decreases the dimension of states and reduce the number of hypotheses by utilizing statistical independence between basic components, while reserves the actual correlation by keeping parallel  LMO densities of random finite subsets. However, we still have a problem that  how to compute the LMO density of each random
subset from the global LMO density, which will be discussed in the following subsection.
 \subsection{Labeled Set Marginal Density}
In  \cite{refr:JMB}, we have given concept of the set marginal density as shown in Definition 3 and its  computing method  for unlabeled RFSs as shown in Lemma 4. In this subsection, Propositions 2$-$5 provide the specified method to compute the set marginal density of the universal LMO density, GLMB density and some special cases of GLMB density including $\delta$-GLMB and M$\delta$-GLMB densities, respectively. The proofs of Propositions 2$-$5 are given in Appendices C$-$F.

\begin{Def} Let $\Psi$ be an RFS. Then for any random finite subset of $\Psi$, denoted by $\Psi_1$, its multi-object density  $\pi_{\Psi_1}(X)$, is defined as the set marginal density of  $\Psi_1$ with respect to $\Psi$.
 \end{Def}
 \begin{Lem}
Let $\Psi$ be an RFS on space $\mathbb{X}$. Then for any random finite subset of $\Psi$, denoted by $\Psi_1$, the set marginal density of $\Psi_1$ with respect to $\Psi$, denoted by $\pi_{\Psi_1}(X)$ can be derived by
\begin{equation}\label{the_marginal_density_of_an_RFS}
 f_{\Psi_1}(X)=\frac{\delta\Pr(\Psi_1\subseteq S,\Psi/\Psi_1\subseteq \mathbb{X})}{\delta X}\bigg{|}_{S=\emptyset}
\end{equation}
where ``$\delta/\delta X$'' denotes a set derivative.
\end{Lem}

\begin{Pro}\label{marginal for LMD}
Assume a labeled RFS $\bPsi$ on state space  $\mathbb{X}\times\mathbb{L}$ and its multi-object density is $\bpi_{\bPsi}(\mathbf{X})=\omega(\mathcal{L}(\mathbf{X}))p(\bX)$.
If $\bPsi_1$ on   $\mathbb{X}\times\mathbb{L}_1$ is a  subset of $\bPsi$  with $\mathbb{L}_1\subseteq\mathbb{L}$, then the labeled set marginal density of $\bPsi_1$ is
 \begin{equation}\label{ the marginal set density of 1}
  \bpi_{\bPsi_1}(\mathbf{X})=\sum_{I\in\mathcal{F}(\mathbb{L}/\mathbb{L}_1)}\omega(\mathcal{L}(\bX)\cup I) p_I(\bX)  \end{equation}
   where
 \begin{equation}\label{set marginal density_where2}
   p_{\{\ell_1,\cdots,\ell_n\}}(\bX)\!=\!\!\int\!\! p(\bX\cup\{(x_1,\ell_1),\!\cdots\!,(x_n,\ell_n)\})dx_1\cdots,dx_n.
 \end{equation}
\end{Pro}
\begin{Pro}
Assume a GLMB RFS $\bPsi$ on state space $\mathbb{X}\times\mathbb{L}$ and its multi-object density has the form of (\ref{GLMB}).
If $\bPsi_1$ on   $\mathbb{X}\times\mathbb{L}_1$ is a  subset of $\bPsi$  with $\mathbb{L}_1\subseteq\mathbb{L}$, then the labeled set marginal
density of $\bPsi_1$ is
 \begin{equation}\label{marginal-GLMB}
  \bpi_{\bPsi_1}(\mathbf{X})=\sum_{I\in\mathcal{F}(\mathbb{L}/\mathbb{L}_1)}\sum_{c\in\mathbb{C}}\omega^{(c)}(\mathcal{L}(\bX)\cup I){[p^{(c)}]}^\bX.
\end{equation}
\end{Pro}
\begin{Pro}
Assume a $\delta$-GLMB RFS $\bPsi$ on state space is $\mathbb{X}\times\mathbb{L}$ and its multi-object density has the same form of (\ref{delta-GLMB}).
If $\bPsi_1$ on   $\mathbb{X}\times\mathbb{L}_1$ is a  subset of $\bPsi$  with $\mathbb{L}_1\subseteq\mathbb{L}$,
then the  labeled set marginal density of $\bPsi_1$ is
 \begin{equation}\label{marginal-dGLMB}
  \bpi_{\bPsi_1}(\mathbf{X})= \sum_{I_2\in\mathcal{F}(\mathbb{L}/\mathbb{L}_1)}\sum_{(I_1,\xi)\in\mathcal{F}(\mathbb{L}_1)\times\Xi}\delta_{I_1}(\mathcal{L}(\bX))\omega^{(I_1\cup I_2,\xi)}{[p^{(\xi)}]}^\bX.
\end{equation}
\end{Pro}

\begin{Pro}
Assume an $M\delta$-GLMB RFS $\bPsi$ on state space is $\mathbb{X}\times\mathbb{L}$ and its multi-object density has the same form of (\ref{Mdelta-GLMB}).
If $\bPsi_1$ on   $\mathbb{X}\times\mathbb{L}_1$ is a  subset of $\bPsi$  with $\mathbb{L}_1\subseteq\mathbb{L}$,
then the labeled set marginal density of $\bPsi_1$ is
 \begin{equation}\label{marginal-dGLMB}
  \bpi_{\bPsi_1}(\mathbf{X})= \sum_{I_2\in\mathcal{F}(\mathbb{L}/\mathbb{L}_1)}\sum_{I_1\in\mathcal{F}(\mathbb{L}_1)}\delta_{I_1}(\mathcal{L}(\bX))\omega^{(I_1\cup I_2)}{[p^{(I_1)}]}^\bX.
\end{equation}
\end{Pro}
\begin{Pro}
Assume an LMB RFS $\bPsi$ on state space is $\mathbb{X}\times\mathbb{L}$ and its multi-object density has the same form of (\ref{LMB}).
If $\bPsi_1$ on   $\mathbb{X}\times\mathbb{L}_1$ is a  subset of $\bPsi$  with $\mathbb{L}_1\subseteq\mathbb{L}$,
then the labeled set marginal density of $\bPsi_1$ is
\begin{align}\label{LMB_M}
\begin{split}  \bpi_{\bPsi_1}(\mathbf{X})=
\sum_{I_2\in\mathcal{F}(\mathbb{L}/\mathbb{L}_1)}\sum_{I_1\in\mathcal{F}(\mathbb{L}_1)}\delta_{I_1}(\mathcal{L}(\bX))\omega^{(I_1\cup I_2)}{p}^\bX.
\end{split}
\end{align}
\end{Pro}
\subsection{Summary}
This section presents an enhanced  approximation approach for LMO density which operates as shown in Fig. 1 at a conceptual level. The proposed approximation approach adopts two-step  strategy: firstly, the actual correlation between different basic components are estimated based on Definition 2, and the basic components are grouping according to the criterion that any two basic components belong to different group  exhibit no correlation; secondly, the labeled set marginal density of each group is computed based on Propositions 2$-$5,  and the original LMO density is decomposed based on Proposition 1.  The proposed approach does not follow the old routine to approximate the LMO density using a certain type of distribution, but decomposes the LMO density  according to the result of correlation analysis, and hence is referred to as correlation analysis (CA) based approximation approach.  The innovation of the proposed approximation is the perception of the actual statistical correlatoion between objects which makes  it possible to utilize the actual independence to decrease the computational complexity as well as reserve the required correlation to obtain high approximation accuracy.

  \begin{figure}[htpb]
\label{tracks_estimate}
  \centering
  \includegraphics[width=6cm]{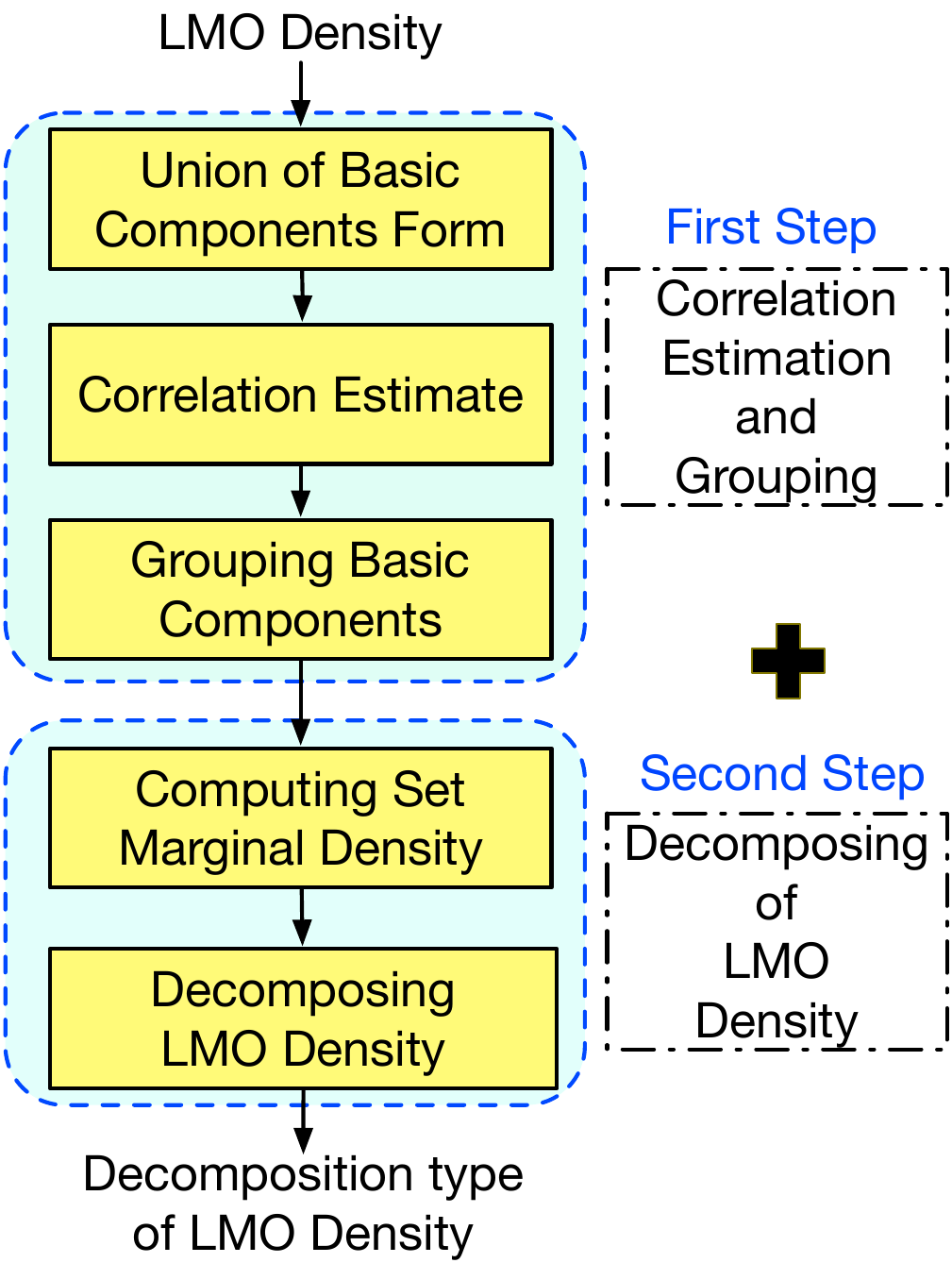}\\
  \caption{The flow diagram of the enhanced approximation approach}\label{approximation_strategy}
\end{figure}

Assume an LMO density $\bpi$ of form (\ref{factorized}) on  state space $\mathbb{X}$ and  label space  $\mathbb{L}$. Here, we compare the $\delta$-GLMB density approximation given in Lemma 2 and the proposed CA based approximation approach. Assume that for the CA  based approximation, the labeled RFS $\bPsi $ are partitioned into $N$ independent subsets, i.e., $\bPsi=\biguplus_{i=1}^N\bPsi_i$ with $\mathbb{L}_i$ the label space of $\bPsi_i$. The total number of hypotheses involving different label sets (except for the empty set) for LMO density and its two approximations are given in Table I, while the number of Euclidean notion of densities with different dimensions are summarized in Table II where $\mathbb{P}=\{\mathbb{L}_1,\cdots,\mathbb{L}_N\}$  and $n_{\text{max}}=\max\limits_{i=1,\cdots,N} |\mathbb{L}_i|$.
\begin{table}[htpb]
\caption{The number of hypotheses}
\begin{center}
\begin{tabular}{c|c}
\hline
\hline

& The number of hypotheses \\
\hline
 LMO density& $2^{|\mathbb{L}|}-1$ \\
\hline
$\delta$-GLMB density approximation& $2^{|\mathbb{L}|}-1$ \\
\hline
CA based approximation & $\sum_{i=1}^N (2^{|\mathbb{L}_i|}-1)$\\
\hline
\end{tabular}
\end{center}
\end{table}\\
\begin{table*}[h]
\caption{The number of Euclidean notion of densities with different dimensions}
\begin{center}
\begin{tabular}{c|c|c|c|c|c|c|c}
\hline
\hline
$\!$&$\mathbb{X}$ & $\cdots$ &   $\mathbb{X}^n$ &    $\cdots$ &$\mathbb{X}^{n_{\text{max}}}$ &$\cdots$ &    $\mathbb{X}^{|\mathbb{L}|}$ \\
\hline
LMO density& $C_{|\mathbb{L}|}^1$  &      $\cdots$   &  $C_{|\mathbb{L}|}^n$     & $\cdots$   &   $C_{|\mathbb{L}|}^{n_{\text{max}}}$&    $\cdots$  &$C_{|\mathbb{L}|}^{|\mathbb{L}|}$\\
\hline
$\delta$-GLMB density approximation& $\sum_{n=1}^{|\mathbb{L}|}n \cdot C_{|\mathbb{L}|}^{n}$   &  0 &  0  &   0 & 0&0&0\\
\hline
CA based approximation & $\sum_{\mathbb{L}_i\in\mathbb{P},|\mathbb{L}_i|\geqslant 1} C_{|\mathbb{L}_i|}^1 $&   $\cdots$    &   $\sum_{\mathbb{L}_i\in\mathbb{P},|\mathbb{L}_i|\geqslant n}   C_{|\mathbb{L}_i|}^n$ &    $\cdots$ &      $C_{n_{\text{max}}}^{n_{\text{max}}}$&   $\cdots$   &  0\\
\hline
\end{tabular}
\end{center}
\end{table*}

It can be seen from Table I that the number of hypotheses for $\delta$-GLMB density approximation is the same as the original LMO density, while the number of hypotheses for CA based approximation is smaller, i.e., $\sum_{i=1}^N(2^{|\mathbb{L}_i|}-1)\leqslant 2^{|\mathbb{L}|}-1$.  It also can been seen that the $\delta$-GLMB density approximation only have densities on single-object state space $\mathbb{X}$, while the CA based on approximation still have high-dimensional densities. However, one should note that the number of densities for   the CA based on approximation is less than that of the LMO density under the  state spaces with the same dimension.  Actually, the reservation of high-dimensional densities for CA based approximation is for the reservation of actual correlation between objects.
\section{Numerical Results}
Consider a labeled RFS $\bPsi$  on space $\mathbb{X}\times\mathbb{L}$, where $\mathbb{X}=\mathbb{R}$ is the field of real number and $\mathbb{L}=\{1,2,3\}$. We design an LMO density of $\bPsi$ shown as
\begin{equation}\label{LMO}
\begin{split}
&\bpi(\bX)=\\
&\left\{
\begin{array}{ll}
0.01, &\bX=\emptyset\\
 0.01\mathcal{N}(x;m_1,R_1), &\bX=\{(x,1)\}\\
 0.01\mathcal{N}(x;m_2,R_2), &\bX=\{(x,2)\}\\
 0.09\mathcal{N}(x;m_3,R_3), &\bX=\{(x,3)\}\\
 0.07\mathcal{N}\left(
 \left(\begin{array}{ll}
 \!\!\!x_1\!\!\!\\
 \!\!\!x_2\!\!\!\\ \end{array}
 \right); \mathbf{m}_{12},\mathbf{R}_{12}\right), &\bX=\{(x_1,1), (x_2,2)\}\\
 0.09\mathcal{N}\left(
 \left(\begin{array}{ll}
 \!\!\!x_1\!\!\!\\
 \!\!\!x_2\!\!\!\\ \end{array}
 \right);\mathbf{m}_{13},\mathbf{R}_{13}\right),&\bX=\{(x_1,1), (x_2,3)\}\\
 0.09\mathcal{N}\left(
 \left(\begin{array}{ll}
 \!\!\!x_1\!\!\!\\
 \!\!\!x_2\!\!\!\\ \end{array}
 \right);\mathbf{m}_{23},\mathbf{R}_{23}\right),&\bX=\{(x_1,2), (x_2,3)\}\\
 0.63\mathcal{N}\left(
 \left(\begin{array}{ll}
 \!\!\!x_1\!\!\!\\
 \!\!\!x_2\!\!\!\\
 \!\!\!x_3\!\!\!\end{array}
 \right);\mathbf{m}_{123},\mathbf{R}_{123}\right), &\bX=\begin{array}{ll}\!\!\left\{(x_1,1),(x_2,2),\right.\\
 \!\!\!\left. (x_3,3)\right\}\end{array}\\
\end{array}\right.
\end{split}
\end{equation}
where
\begin{equation}
\begin{split}
&m_1=1, R_1=1\\
&m_2=2, R_2=2\\
&m_3=8, R_3=3\\
&\mathbf{m}_{12}= \left(\begin{array}{ll}
 \!\!\!1.1\!\!\!\\
 \!\!\!1.2\!\!\!\\ \end{array}
 \right), \mathbf{R}_{12}=\left[\begin{array}{cc}
1.2 &R_0 \\
R_0 &2.2
 \end{array}\right]\\
 &\mathbf{m}_{13}= \left(\begin{array}{ll}
 \!\!\!1\!\!\!\\
 \!\!\!8\!\!\!\\ \end{array}
 \right), \mathbf{R}_{13}=\left[\begin{array}{cc}
1 &0\\
0 &3
 \end{array}\right]\\
 &\mathbf{m}_{23}= \left(\begin{array}{ll}
 \!\!\!2\!\!\!\\
 \!\!\!8\!\!\!\\ \end{array}
 \right), \mathbf{R}_{23}=\left[\begin{array}{cc}
2 &0\\
0 &3
 \end{array}\right]\\
  &\mathbf{m}_{123}= \left(\begin{array}{c}
 \!\!\!1.1\!\!\!\\
 \!\!\!1.2\!\!\!\\
  \!\!\!8\!\!\!\\ \end{array}
 \right), \mathbf{R}_{123}=\left[\begin{array}{ccc}
1.2 &R_0 &0\\
R_0 & 2.2&0\\
0&0& 3\\
 \end{array}\right].\\
\end{split}
\end{equation}
Where $R_0>0$ denotes the covariance of the states with labels $\ell=1$ and $\ell=2$ under the hypothesis $I=\{1,2\}$ or $I=\{1,2,3\}$.
\subsection{Approximations of LMO density}
In the subsection, we give three approximations of the LMO density in (\ref{LMO}).

$\bullet$ $\delta$-GLMB density approximation, $\hat\bpi_{\delta\text{-GLMB}}$, according to Lemma 2;

$\bullet$ CA based approximation, $\hat\bpi_{\text{CA}}$, proposed in Section III;

$\bullet$ CA based approximation of the approximated $\delta$-GLMB density, $\hat\bpi^{\text{CA}}_{\delta\text{-GLMB}}$, which firstly approximates the LMO density as a $\delta$-GLMB density and then
 approximate the resulting $\delta$-GLMB density using the CA based approximation approach.
\subsubsection{$\delta$-GLMB Density Approximation}
 Based on Lemma 2, we approximate (\ref{LMO}) into a $\delta$-GLMB density shown as
 \begin{equation}\label{GLMB-app}
\begin{split}
&\hat\bpi_{\delta\text{-GLMB}}(\bX)=\\
&\left\{
\begin{array}{ll}
0.01, &\bX=\emptyset\\
 0.01\mathcal{N}(x;m_1,R_1), &\bX=\{(x,1)\}\\
 0.01\mathcal{N}(x;m_2,R_2), &\bX=\{(x,2)\}\\
 0.09\mathcal{N}(x;m_3,R_3) &\bX=\{(x,3)\}\\
 0.07\prod\limits_{i\in\{1,2\}}\mathcal{N}(x_i;m_{12}^i,R_{12}^i), &\bX=\{(x_1,1), (x_2,2)\}\\
 0.09\prod\limits_{i\in\{1,2\}}\mathcal{N}(x_i;m_{13}^i,R_{13}^i),&\bX=\{(x_1,1), (x_3,3)\}\\
 0.09\prod\limits_{i\in\{2,3\}}\mathcal{N}(x_i;m_{23}^i,R_{23}^i),&\bX=\{(x_2,2), (x_3,3)\}\\
 0.63\prod\limits_{i\in\{1,2,3\}}\mathcal{N}(x_i;m_{123}^i,R_{123}^i)), &\bX=\begin{array}{ll}\!\!\!\!\!\left\{(x_1,1),(x_2,2),\right.\\
 \!\!\!\left. (x_3,3)\right\}\end{array}\\
\end{array}\right.
\end{split}
\end{equation}
where
\begin{equation}
\begin{split}
&m_{12}^i=\mathbf{m}_{12}(i),R_{12}^i=\mathbf{R}_{12}(i,i), i=1,2\\
&m_{13}^i=\mathbf{m}_{13}(i),R_{13}^i=\mathbf{R}_{13}(i,i), i=1,3\\
&m_{23}^i=\mathbf{m}_{23}(i),R_{23}^i=\mathbf{R}_{23}(i,i), i=2,3\\
&m_{123}^i=\mathbf{m}_{123}(i),R_{123}^i=\mathbf{R}_{123}(i,i,i), i=1,2,3.\\
\end{split}
\end{equation}
\subsubsection{CA based Approximation}
From (\ref{LMO}),  we can extract the distribution of $(E_1,E_2,E_3)$ as Table III,
 \begin{table}[htbp]
\label{table_example_numberical}
\begin{center}
 \caption{The joint distribution of $(E_1, E_2, E_3)$}
 \begin{tabular*}{0.5\textwidth}{@{\extracolsep{\fill}}c|cccc}
\hline
 \diagbox{$E_3$}{$E_1,E_2$} & 00 & 10 &01 &11\\
\hline
0 & $0.01$ & $0.01$ &0.01 &0.07\\
1 & $0.09$ & $0.09$ &0.09 &0.63\\
\hline
 \end{tabular*}
 \end{center}
 \end{table}

According to (\ref{correlation estimate}), we can get the absolute correlation coefficients between each basic components of $\bPsi$, i.e., $\bpsi_1$, $\bpsi_2$ and $\bpsi_3$ as
\begin{equation}
\begin{split}
\label{correlation_12} \gamma_{1,2}=&\frac{\alpha_{1,2}+\beta_{1,2}}{2}\\
=&\frac{0.375+R_0/1.6248}{2}\\
=&0.1875+R_0/3.2856>0.1875\\
\end{split}
\end{equation}
\begin{align}
\label{correlation_23} \gamma_{2,3}=&0\\
\label{correlation_13} \gamma_{1,3}=&0
\end{align}
where $\omega_E=\omega_S=\frac{1}{2}$ in (\ref{correlation estimate}).
Hence, we can conclude that $\bpsi_3$  is independent of $\bpsi_1$ and $\bpsi_2$, and $\bpsi_1$ and $\bpsi_2$ do have correlation.
We can divide $\bPsi$ into two independent subsets, namely, $\bpsi_1\cup\bpsi_2$ and $\bpsi_3$.

Let $\bPsi_{a}=\bpsi_1\cup\bpsi_2$ and $\bPsi_{b}=\bpsi_3$. According to Proposition 2, we can compute the labeled  set marginal density of $\bPsi_a$ and $\bPsi_b$ as
\begin{equation}\label{MLMD1-example-2}
\begin{split}
&\bpi_{\bPsi_a}=\\
&\left\{\begin{array}{ll}
\ 0.1, &\bX=\emptyset\\
\ 0.1\mathcal{N}(x;m_{a,1},R_{a,1}), &\bX=\{(x,1)\}\\
\ 0.1\mathcal{N}(x;m_{a,2},R_{a,2}), &\bX=\{(x,2)\}\\
\ 0.7\mathcal{N}\left(
 \left(\begin{array}{ll}
 \!\!\!x_1\!\!\!\\
 \!\!\!x_2\!\!\!\\ \end{array}
 \right);\mathbf{m}_{a,12},\mathbf{R}_{a,12}\right), &\bX=\{(x_1,1),(x_2,2)\}
\end{array}\right.
\end{split}
\end{equation}
where
\begin{equation}
\begin{split}
&m_{a,1}=1, R_{a,1}=1\\
&m_{a,2}=2,R_{a,2}=2\\
&\mathbf{m}_{a,12}=\left(\begin{array}{ll}
 \!\!\!1.1\!\!\!\\
 \!\!\!1.2\!\!\!\\ \end{array}
 \right),\mathbf{R}_{a,12}=\left[\begin{array}{cc}
1.2&1\\
1&2.2
 \end{array}\right]
 \end{split}
 \end{equation}
 and
\begin{equation}
\bpi_{\bPsi_b}(\bX)=\left\{\begin{array}{ll}
\ 0.1, &\bX=\emptyset\\
\ 0.9\mathcal{N}(x;m_{b,3},R_{b,3}), &\bX=\{(x,3)\}
\end{array}\right.
\end{equation}
where
\begin{equation}
m_{b,3}=3, R_{b,3}=8.
\end{equation}
Finally, we can obtain the CA based approximation as
\begin{align}\label{SMD-approx-example}
\hat\bpi_{\text{CA}}(\bX)&=\bpi_{{\bPsi}_a}(\bX\cap\mathbb{X}\times\mathbb{L}_a) \bpi_{{\bPsi}_b}(\bX\cap\mathbb{X}\times\mathbb{L}_b)
\end{align}
with $\mathbb{L}_a=\{1,2\}$ and $\mathbb{L}_b=\{3\}$.
\subsubsection{CA based Approximation of  the Approximate GLMB
Density}

Let $\hat\bPsi$ denotes the approximate $\delta$-GLMB RFS whose density is (\ref{GLMB-app}), and $\hat\bpsi_\ell$s, $\ell\in\{1,2,3\}$ denote the basic components of $\hat\bPsi$.

According to (\ref{correlation estimate}), we can get the absolute correlation coefficients between $\hat\bpsi_1$, $\hat\bpsi_2$ and $\hat\bpsi_3$  as
\begin{align}
\label{correlation_12G}\gamma_{1,2}=&\frac{\alpha_{1,2}}{2}=0.1875\\
\label{correlation_23G}\gamma_{2,3}=&0\\
\label{correlation_13G}\gamma_{1,3}=&0
\end{align}
where $\omega_E=\omega_S=\frac{1}{2}$ in (\ref{correlation estimate}).   One can find that the approximate GLMB  density does lose a part of correlation towards the original LMO density comparing (\ref{correlation_12}) and (\ref{correlation_12G}).

Hence, we can also conclude that $\hat\bpsi_3$  is independent of both $\hat\bpsi_1$ and $\hat\bpsi_2$, and $\hat\bpsi_1$ and $\hat\bpsi_2$ do have correlation.
Then we can also divide $\hat\bPsi$ into two independent subsets, namely, $\hat\bpsi_1\cup\hat\bpsi_2$ and $\hat\bpsi_3$.

Let $\hat\bPsi_a=\hat\bpsi_1\cup\hat\bpsi_2$ and $\bPsi_b=\hat\bpsi_3$. According to Proposition 3, we can compute the labeled set marginal density of $\hat\bPsi_a$ and $\hat\bPsi_b$ as
 \begin{align}\label{GLMB1-SMD-example}
  &\bpi_{\hat\bPsi_a}(\mathbf{X})\\
 &=\left\{
  \begin{array}{ll}
 0.1, &\bX=\emptyset\\
   0.1\mathcal{N}(x;\hat m_{a,1},\hat R_{a,1}),&\bX=(x,1)\\
   0.1\mathcal{N}(x;\hat m_{a,2},\hat R_{a,2}),&\bX=(x,2)\\
   0.7\!\!\prod \limits_{i\in\{1,2\}}\!\!\mathcal{N}((x_i;\hat m_{a,12}^i,\hat R_{a,12}^i), &\bX=\{(x_1,1),(x_2,2)\}
   \end{array}\right.
\end{align}
where
\begin{equation}
\begin{split}
&\hat m_{a,1}=1, \hat R_{a,1}=1\\
&\hat m_{a,2}=2,\hat R_{a,2}=2\\
&\hat m^1_{a,12}=1.1, \hat R^1_{a,12}=1.2\\
&\hat m^2_{a,12}=2.2, \hat R^2_{a,12}=2.2\\
\end{split}
\end{equation}

 \begin{align}\label{GLMB1-SMD-example}
  &\bpi_{{\hat\bPsi}_b}(\mathbf{X})\left\{
  \begin{array}{ll}
 \ 0.1, &\bX=\emptyset\\
  \ 0.9\mathcal{N}(x;\hat m_{b,3},\hat m_{b,3}),&\bX=(x,3)\\
    \end{array}\right.
\end{align}
where
\begin{equation}
\begin{split}
&\hat m_{b,3}=3, \hat R_{b,3}=8.
\end{split}
\end{equation}

Hence, the CA based approximation of  the approximated GLMB
density is
\begin{align}\label{SMD-approx-example}
\hat \bpi_{\delta\text{-GLMB}}^{\text{CA}}(\bX)&=\bpi_{{\hat\bPsi}_a}(\bX\cap\mathbb{X}\times\mathbb{L}_a) \hat\bpi_{\hat{\bPsi}_b}(\bX\cap\mathbb{X}\times\mathbb{L}_b)
\end{align}
with $\mathbb{L}_a=\{1,2\}$ and $\mathbb{L}_b=\{3\}$.

\subsection{Computational Complexity and Approximate Error Analysis}
\subsubsection{Computational Complexity}
In order to evaluate the computational complexity of the original LMO density and its approximations, some important quantities of $\bpi$, $\hat\bpi_{\delta \text{-GLMB}}$, $\hat\bpi_{\text{CA}}$ and $\hat\bpi_{\delta\text{-GLMB}}^{\text{CA}}$ are summarized as Table II,
 \begin{table}[htbp]
\label{table_example_numberical}
\begin{center}
 \caption{Computation Analysis}
 \begin{tabular*}{0.5\textwidth}{@{\extracolsep{\fill}}|c|c|c|c|c|c|}
 \hline
 \,  & $T_0$& $T_1$  &  $T_2$  &$T_3$     &Correlation loss\\
\hline
$\bpi$          &  8   &      3   & 3  &1 & \,\\
\hline
$\hat\bpi_{\delta\text{-GLMB}}$ & 8     &      12    & 0 &0 & yes\\
\hline
$\hat\bpi_{\text{CA}}$  &4    &          3 & 1 &0 & no \\
\hline
$\hat\bpi^{\text{CA}}_{\delta\text{-GLMB}}$ &4    & 5 & 0 &0 & yes (same as $\hat\bpi_{\delta\text{-GLMB}}$)\\
\hline
 \end{tabular*}
 \end{center}
 \end{table}

$T_0$: NO. of hypotheses

$T_1$: NO. of densities on $\mathbb{X}$

$T_2$: NO. of densities on $\mathbb{X}^{2}$

$T_3$: NO. of densities on $\mathbb{X}^{3}$

Note that the hypotheses here involves different existing target label sets $I$, and the hypothesis $I=\emptyset$ is omitted because it can be determined by other hypotheses totally.
\subsubsection{Approximate Error}
Herein, we evaluate the approximate error of three approximations in terms of KLD for different values of orgrinal $\beta_{1,2}$. $\beta_{1,2}$ is the absolute correlation coefficient between the states of basic components $\bpsi_1$ and $\bpsi_2$. The value  of $\beta_{1,2}$ ranges from 0 to 1.  In this example, the relationship between the quantities $R_0$ and $\beta_{1,2}$ is that  $\beta_{1,2}=|\sqrt{1.2\times2.2}\times R_0|\cong |1.625R_0|$.

  \begin{figure}[htpb]
\label{tracks_estimate}
  \centering
  \includegraphics[width=8cm]{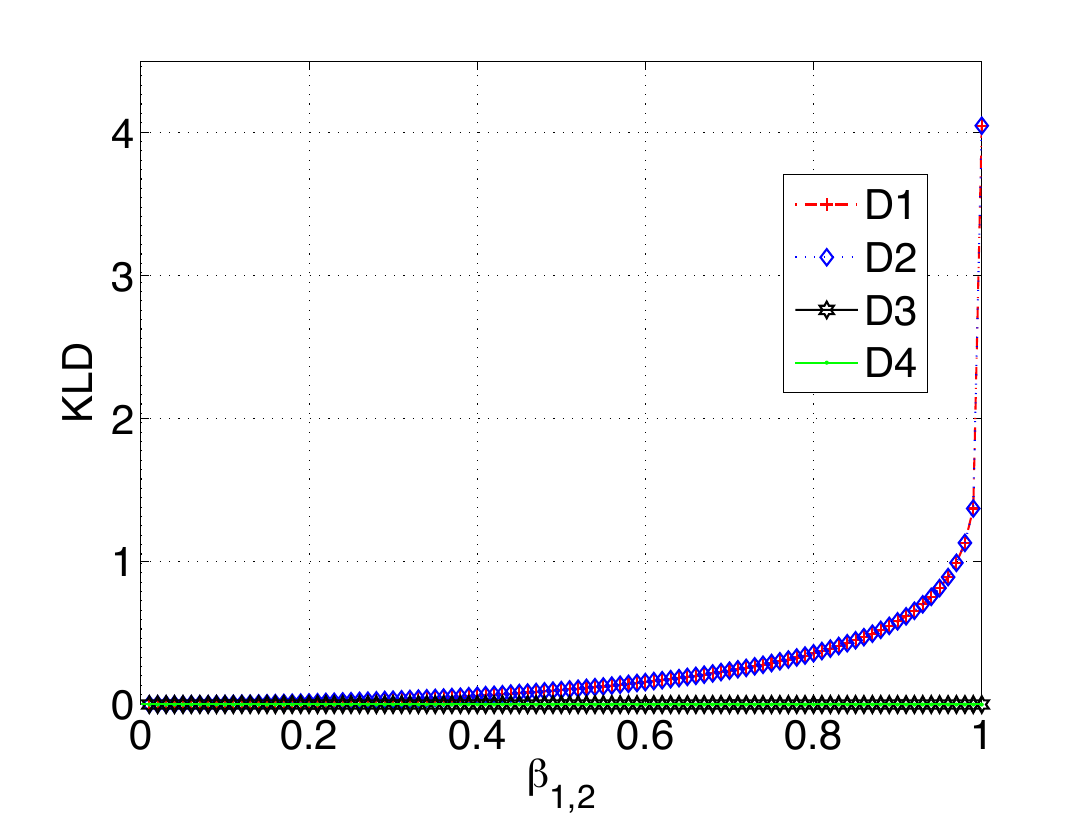}\\
  \caption{the curves of the KLDs: D1, D2, D3 and   D4 vs $\beta_{1,2}$, where D1 denotes $D_{\text{KL}}(\bpi;\hat\bpi_{\delta\text{-GLMB}})$, D2 denotes $D_{\text{KL}}(\bpi;\hat\bpi_{\delta\text{-GLMB}}^{\text{CA}})$, D3 denotes $D_{\text{KL}}(\bpi;\hat\bpi_{\text{CA}})$ and D4 denotes $D_{\text{KL}}(\hat\bpi_{\delta\text{-GLMB}};\hat\bpi^{\text{CA}}_{\delta\text{-GLMB}})$.}\label{Marginal_arameters}
\end{figure}

The KLD between $\hat\bpi_{\text{CA}}$ and $\bpi$, $\hat\bpi_{\delta\text{-GLMB}}$ and $\bpi$, $\hat\bpi^{\text{CA}}_{\delta\text{-GLMB}}$ and $\bpi$ is computed respectively according to (\ref{KLD}).  The KLD between  $\hat\bpi_{\delta\text{-GLMB}}$ and $\hat\bpi^{\text{CA}}_{\delta\text{-GLMB}}$ also is computed. Fig. 1 shows the curves of the KLDs: $D_{\text{KL}}(\bpi;\hat\bpi_{\delta\text{-GLMB}})$, $D_{\text{KL}}(\bpi;\hat\bpi_{\text{CA}})$, $D_{\text{KL}}(\bpi;\hat\bpi^{\text{CA}}_{\delta\text{-GLMB}})$ and   $D_{\text{KL}}(\bpi_{\delta\text{-GLMB}});\hat\bpi^{\text{CA}}_{\delta\text{-GLMB}})$ vs $\beta_{1,2}$.
\subsubsection{Summary}
$\delta$-GLMB approximation of LMO density actually approximates the conditional joint probability density $p(\bX)$ under each hypothesis as the product of its marginal densities, as a result it totally discards the correlation between states of  basic components, no matter how strong the  correlation of the original LMO density is.  Hence, the stronger the original correlation is, the larger the approximate error is, which can be reflected from the curve of D1 in Fig. 2. The $\delta$-GLMB density approximation also retains all hypotheses.
As shown in Table II, the number of hypotheses of $\hat\bpi_{\delta\text{-GLMB}}$ are the same as $\bpi$, and all the densities of $\hat\bpi_{\delta\text{-GLMB}}$  are on the single-object
space $\mathbb{X}$. Further comparing $\hat\bpi_{\delta\text{-GLMB}}^{\text{CA}}$
with $\hat\bpi_{\delta\text{-GLMB}}$, we find that $\hat\bpi_{\delta\text{-GLMB}}$ has redundant statistical information, for it can be further simplified by reducing the number of hypotheses
 and number of densities without approximate error,  as shown in Table IV and Fig. 2.

As for $\hat\bpi_{\text{CA}}$, it reduces the number of hypotheses by utilizing independence, and also retains high-dimensional densities to keep correlation. As shown in Table II,
even though $\hat\bpi_{\text{CA}}$ has a density on space $\mathbb{X}^2$ while $\hat\bpi_{\delta\text{-GLMB}}$
does not have, $\hat\bpi_{\text{CA}}$ only has 3 densities on space $\mathbb{X}$ while $\hat\bpi_{\delta\text{-GLMB}}$ has 12.
Furthermore, the high-dimensional density of $\hat\bpi_{\text{CA}}$ is retained in return for keeping required correlation. As shown in Fig. 1, the curve of D3 reflects that the approximate error form the $\hat \bpi_{\text{CA}}$ to the full LMO density $\bpi$ keeps zero for different values of $\beta_{1,2}$. Hence, $\hat\bpi_{\text{CA}}$ is a kind of approximation which can balance the computational complexity and approximate error.
\section{Conclusion}
In this paper, we proposed an enhanced approximation of labeled multi-object (LMO)
 density which  evaluates the correlation between objects adaptively and factorizes the LMO density into  densities of several independent subsets according to the
 correlation analysis. Furthermore, to obtain a tractable factorization of LMO density, we derived the labled  set marginal density of any subset of the universal labeled RFS, and GLMB RFS family and its subclasses.
Unlike the conventional approximate approach which sacrifices statistical correlation  for computational efficiency, the proposed method takes into account the simplification
of the complicate structure of  LMO density and the reservation of necessary correlation at the same time.
\appendices
\section{Proof of Equation (22)}
\begin{proof}
For an arbitrary labeled RFS $\bPsi$, its LMO density has the form  of (\ref{factorized}). Based on Definition 1, $\bPsi$ can be represented as the union of basic components, i.e., $\bPsi=\cup_{\ell\in\mathbb{L}} \bpsi_\ell$. To describe the uncertainty existence of the basic component, we define a random variable $E_\ell$ for each basic component $\bpsi_\ell$. The statistics of all $E_\ell$s, $\ell\in\mathbb{L}$ are distributed according to the joint probability distribution given in (\ref{distribution of existence}).

For any $\ell\neq\ell'\in\mathbb{L}$, to compute the correlation coefficient between $E_\ell$ and $E_{\ell'}$, the  marginal  probability distribution of $E_\ell$ and $E_{\ell'}$ should be computed  firstly from (\ref{distribution of existence}), and is given in Table V
 \begin{table}[htbp]
\label{table_example}
\begin{center}
 \caption{The Joint Distributions of $E_\ell$ and $E_{\ell'}$}
 \begin{tabular*}{0.3\textwidth}{@{\extracolsep{\fill}}c|cc}
\hline
 \diagbox{$E_{\ell'}$}{$E_{\ell}$} & 0 & 1 \\
\hline
0 & $\epsilon_{\ell,\ell'}^{(0,0)}$ & $\epsilon_{\ell,\ell'}^{(1,0)}$ \\
1 & $\epsilon_{\ell,\ell'}^{(0,1)}$ & $\epsilon_{\ell,\ell'}^{(1,1)}$\\
\hline
 \end{tabular*}
 \end{center}
 \end{table}
where  $\epsilon^{(i,j)}_{\ell,\ell'}$ is given in (\ref{Ma_probability_E}).

According to the joint distribution of $E_\ell$ and $E_{\ell'}$, we can compute the means and variances of $E_{\ell}$, $E_{\ell'}$ as, respectively,
\begin{equation}\label{mean and variane}
\begin{split}
  m_{E_\ell}=&\epsilon_{\ell,\ell'}^{(1,0)}+\epsilon_{\ell,\ell'}^{(1,1)}\\
  m_{E_{\ell'}}=&\epsilon_{\ell,\ell'}^{(0,1)}+\epsilon_{\ell,\ell'}^{(1,1)}\\
  \sigma ^2_{E_\ell}=&(1-\epsilon_{\ell,\ell'}^{(1,0)}-\epsilon_{\ell,\ell'}^{(1,1)})(\epsilon_{\ell,\ell'}^{(1,0)}+\epsilon_{\ell,\ell'}^{(1,1)})\\
   \sigma^2_{E_{\ell'}}=&(1-\epsilon_{\ell,\ell'}^{(0,1)}-\epsilon_{\ell,\ell'}^{(1,1)})(\epsilon_{\ell,\ell'}^{(0,1)}+\epsilon_{\ell,\ell'}^{(1,1)})\\
  \end{split}
\end{equation}
Also, the covariance of $E_\ell$ and $E_{\ell}$ can be computed by
\begin{equation}\label{covariance}
\begin{split}
  \mbox{cov}(E_\ell,E_{\ell})=&m_{E_{\ell}E_{\ell'}}-m_{E_\ell}m_{E_{\ell'}}\\
  =&\epsilon_{\ell,\ell'}^{(1,1)}-(\epsilon_{\ell,\ell'}^{(10)}+\epsilon_{\ell,\ell'}^{(1,1)})(\epsilon_{\ell,\ell'}^{(0,1)}+\epsilon_{\ell,\ell'}^{(1,1)})
\end{split}
\end{equation}
Thus based on the definition of correlation coefficient in (\ref{correlation coefficient}), $\rho_{E_\ell,E_{\ell'}}$ can be computed as
\begin{small}
\begin{equation}\label{correlation_coefficient}
\begin{split}
&\rho_{E_\ell,E_{\ell'}}=\frac{ \text{cov}(E_\ell,E_{\ell'})}{\sqrt{{\sigma^2_{E_\ell}}{\sigma^2_{E_{\ell'}}}}}=\\
&\frac{\epsilon_{\ell,\ell'}^{(1,1)}-(\epsilon_{\ell,\ell'}^{(1,0)}+(\epsilon_{\ell,\ell'}^{(1,1)})(\epsilon_{\ell,\ell'}^{(0,1)}+(\epsilon_{\ell,\ell'}^{(1,1)})}{\sqrt{(\epsilon_{\ell,\ell'}^{(1,0)}\!+\!\epsilon_{\ell,\ell'}^{(1,1)})(1\!-\!\epsilon_{\ell,\ell'}^{(1,0)}\!-\!\epsilon_{\ell,\ell'}^{(1,1)})(\epsilon_{\ell,\ell'}^{(0,1)}\!+\!\epsilon_{\ell,\ell'}^{(1,1)})(1\!-\!\epsilon_{\ell,\ell'}^{(0,1)}\!-\!\epsilon_{\ell,\ell'}^{(1,1)})}}\end{split}
\end{equation}
\end{small}
\end{proof}
\section{Proof of Proposition 1}
\begin{proof}
If a labeled RFS $\bPsi$ on space $\mathbb{X}\times\mathbb{L}$ can be divided into $N$ independent labeled random subsets
$\bPsi_i$ on space $\mathbb{X}\times\mathbb{L}_i, i=1,\cdots, N$, i.e., $ \bPsi=\bigcup_{i=1}^N  \bPsi_i$ with $\mathbb{L}=\mathbb{L}_1\uplus\cdots\uplus\mathbb{L}_N$,
 then according to Lemma 3,  the probability density of $\bPsi$  is related to
the probability densities of $\bPsi_1,\cdots,\bPsi_n$ as
\begin{equation}\label{group-label}
\bpi_\bPsi(\bX)=\sum_{\bW_1\uplus\cdots\uplus \bW_n=\bX} \bpi_{\bPsi_1}(\bW_1)\cdots \bpi_{\bPsi_n}(\bW_n).
\end{equation}
Consider the subset of $\bX$, i.e., $\bX\cap\mathbb{X}\times\mathbb{L}_i=\{\bx\in\bX:\mathcal{L}(\bx)\in\mathbb{L}_i\}$. For any $\bW_i$, $i=1,\cdots,N$, if $\bW_i\nsubseteq\bX\cap\mathbb{X}\times\mathbb{L}_i$, then $\bpi_{\bPsi_i}(\bW_i)=0$.  Furthermore, due to the constrain of $\bW_1\uplus\cdots\uplus \bW_n=\bX$, $\bpi_{\bPsi_1}(\bW_1)\cdots \bpi_{\bPsi_n}(\bW_n)\neq 0$ when and only when each $\bW_i=\bX\cap\mathbb{X}\times\mathbb{L}_i$, $i=1,\cdots,N$. Hence, only one  item in the sum of (\ref{group-label}) works, i.e.,
\begin{equation}\label{SMD-approx-2}
\bpi_{{\bPsi}}(\bX)=\bpi_{{\bPsi}_1}(\bX\cap\mathbb{X}\times\mathbb{L}_1)\cdots \bpi_{{\bPsi}_N}(\bX\cap\mathbb{X}\times\mathbb{L}_N).
\end{equation}
\end{proof}
\section{Proof of Proposition 2}
\begin{proof}
 According to the Definition 3, the set marginal density of $\bPsi_1$ is precisely the LMO density of $\bPsi_1$. Here, our aim is to compute the LMO density of $\bPsi_1$ from the LMO density of $\bPsi$, i.e., $\bpi_{\bPsi}(\bX)=\omega(\mathcal{L}(\bX))p(\bX)$. We will firstly construct the belief-mass function of $\bPsi_1$ using the statistical parameters of  $\bPsi$, i.e., $\{\omega(\mathcal{L}(\bX)),p(\bX)\}_{\mathcal{L}(\bX)\subseteq\mathbb{L}}$, and then derive the LMO density of $\bPsi_1$ via set derivative.

The belief-mass function  of $\bPsi_1$ \cite{refr:Mahler_book}  is,
\begin{equation}\label{Pro2-proof-0}
\beta_{\bPsi_1}(S\times\Pi)
=\Pr(\bPsi_1\subseteq S\times\Pi)
\end{equation}
where $S\subseteq\mathbb{X}$ and $\Pi\subseteq\mathbb{L}_1$ with $\mathbb{L}_1$ the label space of $\bPsi_1$.

Let $\bPsi_2=\bPsi/\bPsi_1$. $\bPsi_2$ is also a random finite  subset of $\bPsi$ and the state space of $\bPsi_2$ is $\mathbb{X}\times\mathbb{L}_2$, where $\mathbb{L}_2=\mathbb{L}/\mathbb{L}_1$. Thus we have
\begin{equation}
\Pr(\bPsi_2\subseteq\mathbb{X}\times\mathbb{L}_2)=1.
\end{equation}
Hence, (\ref{Pro2-proof-0}) can be further represented as
\begin{equation}\label{Pro2-proof-1}
\begin{split}
&\beta_{\bPsi_1}(S\times\Pi)\\
=&\Pr(\bPsi_1\subseteq S\times\Pi,\bPsi_2\subseteq\mathbb{X}\times\mathbb{L}_2)\\
=&\sum_{I_1\subseteq\Pi}\sum_{I_2\in\mathcal{F}(\mathbb{L}_2)}\!\!\Pr(\bPsi_1\subseteq S\times\Pi,\bPsi_2\subseteq\mathbb{X}\times\mathbb{L}_2|\\
&L_1=I_1,L_2=I_2)\Pr(L_1=I_1,L_2=I_2)\\
\end{split}
\end{equation}
where $L_i=\mathcal{L}(\bPsi_i)$ is the set of labels of $\bPsi_i$, $i=1,2$.

Observing (\ref{Pro2-proof-1}), one can find that the probability $\Pr(\bPsi_1\subseteq S\times\Pi, \bPsi_2\subseteq\mathbb{X}\times\mathbb{L}_2|L_1=I_1,L_2=I_2)$ and $\Pr(L_1=I_1,L_2=I_2)$ can be constructed by the statistical parameters $p(\bX)$ and $\omega(\mathcal{L}(\bX))$ respectively.

As $p(\bX)$ is the joint probability density  on $\mathbb{X}$ conditional on the set of labels  $\mathcal{L}(\bX)$,  based on Radon-Nikodym Theorems, the probability $\Pr(\bPsi_1\subseteq S\times\Pi, \bPsi_2\subseteq\mathbb{X}\times\mathbb{L}_2|L_1=I_1,L_2=I_2)$ in (\ref{Pro2-proof-1}) is related to $p(\bX)$ ($\mathcal{L}(\bX)=I_1\cup I_2$) as
\begin{equation}
\begin{split}
&\Pr(\bPsi_1\subseteq S\times\Pi,\bPsi_2\subseteq\mathbb{X}\times\mathbb{L}_2|L_1=I_1,L_2=I_2)=\\
&
\int_{S^{|I_1|}}\int_{\mathbb{X}^{|I_2|}}p(\{(x_{1},I^v_1(1)),\cdots,(x_{n_1},I^v_1(n_1)),(x'_{1},I^v_2(1)),\\
&\cdots,(x'_{n_2},I^v_2(n_2))\})
dx_{1}\cdots dx_{n_1}dx'_{1}\cdots dx'_{n_2}
\end{split}
\end{equation}
where $n_1=|I_1|$, $n_2=|I_2|$, and $I_i^v$ denotes the vector constructed by stacking the elements of $I_i$ in a certain order, $i=1,2$.

Also as $\omega(I)$ denotes the joint existence probability of the label set $I$, the probability $\Pr(L_1=I_1,L_2=I_2)$ in (\ref{Pro2-proof-1}) is related to $\omega(I_1\cup I_2)$ as
\begin{equation}
\Pr(L_1=I_1,L_2=I_2)=\Pr(L=I_1\cup I_2)=\omega(I_1\cup I_2)
\end{equation}
where $L$ denotes the set of labels of $\bPsi$.
Thus, (\ref{Pro2-proof-1}) can be computed by
\begin{equation}\label{Pro2-proof-2}
\begin{split}
&\beta_{\bPsi_1}(S\times\Pi)\\
=&\!\!\sum_{I_1\subseteq\Pi}\sum_{I_2\in\mathcal{F}(\mathbb{L}_2)}\!\!\!\omega(I_1\!\cup\! I_2)\!\int_{S^{|I_1|}}\!\int_{\mathbb{X}^{|I_2|}}\!p(\left\{(x_{1},I^v_1(1)),\!\cdots\!,\! (x_{n_1},I^v_1(n_1)),\right.\\
&\left.,(x'_{1},I^v_2(1)),\cdots,(x'_{n_2},I^v_2(n_2))\right\})
dx_{1}\cdots dx_{n_1}dx'_{1}\cdots dx'_{n_2}\\
=&\sum_{I_1\subseteq\Pi}\!\int_{S^{|I_1|}}\!\!\sum_{I_2\in\mathcal{F}(\mathbb{L}_2)}\!\!\omega(I_1\!\cup\! I_2)\int_{\mathbb{X}^{|I_2|}}\!p(\left\{(x_{1},I^v_1(1)),\!\cdots\!,(x_{n_1},I^v_1(n_1)),\right.\\
&\left.(x'_{1},I^v_2(1)),\cdots,(x'_{n_2},I^v_2(n_2))\right\})
dx'_{1}\cdots dx'_{n_2}dx_{1}\cdots dx_{n_1}.
\end{split}
\end{equation}

Let
\begin{equation}\label{set-marginal-density}
\begin{split}
&\bpi_{\bPsi_1}(\{(x_1,\ell_1),\cdots,(x_{n_1},\ell_{n_1})\})=\\
&\sum_{I_2\in\mathcal{F}(\mathbb{L}_2)}\omega(\{\ell_1,\cdots,\ell_{n_1}\}\cup I_2)\int_{\mathbb{X}^{|I_2|}}p(\left\{(x_{1},\ell_1),\cdots,(x_{n_1},\ell_{n_1}),\right.\\
&\left.(x'_{1},I^v_2(1)),\cdots,(x'_{n_2},I^v_2(n_2))\right\})
dx'_{1}\cdots dx'_{n_2}.
\end{split}
\end{equation}
Then (\ref{Pro2-proof-2}) is further represented as
\begin{equation}\label{Pro2-proof-3}
\begin{split}
&\beta_{\bPsi_1}(S\times\Pi)\\
=&\sum_{n_1=0}^{|\mathbb{L}_1|}\frac{1}{n_1!}\sum_{(\ell_1,\cdots,\ell_{n_1})\in\Pi^{n_1}}\int_{S^{n_1}}\bpi_{\bPsi_1}(\{(x_1,\ell_1),\cdots,(x_{n_1},\ell_{n_1})\})\\
&dx_1,\cdots,dx_{n_1}\\
=&\int_{S\times\Pi}\bpi_{\bPsi_1}(\bX)\delta\bX.
\end{split}
\end{equation}

According to Lemma 4, taking set derivative (on labeled object state space) of $\beta_{\bPsi_1}(S\times\Pi)$ in (\ref{Pro2-proof-3}) and setting $S\times\Pi=\emptyset$, we find that the  set marginal density of $\bPsi _1$ on space $\mathbb{X}\times\mathbb{L}_1$  is precisely $\bpi_{\bPsi_1}(\cdot)$ shown in (\ref{set-marginal-density}).

Let
 \begin{equation}
 \begin{split}
 \label{set marginal density_where2}
   p_{\{\ell_1,\cdots,\ell_n\}}&(\bX)=\\
   &\int_{\mathbb{X}^n} p(\bX\cup\{(x_1,\ell_1),\cdots,(x_n,\ell_n)\})dx_1\cdots,dx_n,
   \end{split}
 \end{equation}
$\bpi_{\bPsi_1}(\bX)$ in (\ref{set-marginal-density}) can be further represented as
\begin{equation}\label{ the marginal set density of 2}
\begin{split}
  \bpi_{\bPsi_1}(\mathbf{X})=&\sum_{I\in\mathcal{F}(\mathbb{L}/\mathbb{L}_1)}\omega(\mathcal{L}(\bX)\cup I) p_{I}(\bX).
\end{split}
\end{equation}
Hence, the proposition holds.
\end{proof}
\section{Proof of Proposition 3}

\begin{proof}
Based on Lemma 1, any LMO density can be factorized as the form of (\ref{factorized}), thus the GLMB density of form (\ref{GLMB}) can be factorized as
\begin{equation}\label{GLMB_factorized}
  \bpi_{\text{GLMB}}(\mathbf{X})=\omega(\mathcal{L(\bX)})p(\bX)
\end{equation}
where
\begin{equation}\label{GLMB_factorized}
\begin{split}
  \omega(\mathcal{L}(\bX))&=\sum_{c\in\mathbb{C}}\omega^{(c)}(\mathcal{L}(\bX))\\
  p(\bX)&=\frac{1}{\sum_{c\in\mathbb{C}}\omega^{(c)}(\mathcal{L}(\bX))}\sum_{c\in\mathbb{C}}\omega^{(c)}(\mathcal{L}(\bX)){[p^{(c)}]}^\bX\\
  \end{split}
\end{equation}
Based on Proposition 2, to compute the set marginal density of $\bPsi_1$, we firstly compute
 \begin{equation}\label{set marginal density_where2}
 \begin{split}
   &p_{\{\ell_1,\cdots,\ell_n\}}(\bX)\\
   =&\int p(\bX\cup\{(x_1,\ell_1),\cdots,(x_2,\ell_n)\})dx_1\cdots,dx_n\\
   =&\frac{\sum_{c\in\mathbb{C}}\omega^{(c)}(\mathcal{L}(\bX)\cup \{\ell_1,\cdots,\ell_n\}){[p^{(c)}]}^\bX}{\sum_{c\in\mathbb{C}}\omega^{(c)}(\mathcal{L}(\bX)\cup \{\ell_1,\cdots,\ell_n\})}. \end{split}
 \end{equation}
Hence, the set marginal density of $\bPsi_1$ in (\ref{ the marginal set density of 2}) can be computed by\begin{equation}
\begin{split}
 &\bpi_{\bPsi_1}(\mathbf{X})\\
 =&\sum_{I\in\mathcal{F}(\mathbb{L}/\mathbb{L}_1)}\sum_{c\in\mathbb{C}}\omega^{(c)}(\mathcal{L}(\bX)\cup I)\frac{\sum_{c\in\mathbb{C}}\omega^{(c)}(\mathcal{L}(\bX)\cup I){[p^{(c)}]}^\bX }{\sum_{c\in\mathbb{C}}\omega^{(c)}(\mathcal{L}(\bX)\cup I)} \\
 =&\sum_{I\in\mathcal{F}(\mathbb{L}/\mathbb{L}_1)}\sum_{c\in\mathbb{C}}\omega^{(c)}(\mathcal{L}(\bX)\cup I){[p^{(c)}]}^\bX
\end{split}
\end{equation}
\end{proof}

\section{Proof of Proposition 4}
\begin{proof}
A $\delta$-GLMB density is a special case of  GLMB density with
\begin{equation}\label{delta-GLMB-where}
\begin{split}
\mathbb{C}=&\mathcal{F}(\mathbb{L})\times\Xi\\
\omega^{(c)}(L)=&\omega^{(I,\xi)}(L)=\omega^{(I,\xi)}\delta_I(L)\\
p^{(c)}=&p^{(I,\xi)}=p^{(\xi)}.
\end{split},
\end{equation}
Hence, according to Proposition 3, the set marginal density of $\bPsi_1$ with respect to the $\delta$-GLMB RFS $\bPsi$ can be obtained  by substitution of (\ref{delta-GLMB-where}) into (\ref{marginal-GLMB}), i.e.,
\begin{equation}\label{Pro4-proof-1}
\begin{split}
 & \bpi_{\bPsi_1}(\mathbf{X})\\
  =&\sum_{I_2\in\mathcal{F}(\mathbb{L}/\mathbb{L}_1)}\sum_{(I,\xi)\in\mathcal{F}(\mathbb{L})\times\Xi}\delta_{I}(\mathcal{L}(\bX)\cup I_2)\omega^{(I,\xi)}{[p^{(\xi)}]}^\bX.\\=&\sum_{I_2\in\mathcal{F}(\mathbb{L}/\mathbb{L}_1)}\sum_{(I_1,\xi)\in\mathcal{F}(\mathbb{L}_1)\times\Xi}\delta_{I_1}(\mathcal{L}(\bX))\omega^{(I_1\cup I_2,\xi)}{[p^{(\xi)}]}^\bX.
\end{split}
\end{equation}
\end{proof}

\section{Proof of Proposition 5}
\begin{proof}
An M$\delta$-GLMB density is a special case of  GLMB density with
\begin{equation}\label{Mdelta-GLMB-where}
\begin{split}
\mathbb{C}=&\mathcal{F}(\mathbb{L})\\
\omega^{(c)}(L)=&\omega^{(I)}(L)=\omega^{(I)}\delta_I(L)\\
p^{(c)}=&p^{(I)}.
\end{split},
\end{equation}
Hence, according to Proposition 3, the set marginal density of $\bPsi_1$ with respect to the M$\delta$-GLMB RFS $\bPsi$ can be obtained  by substitution of (\ref{Mdelta-GLMB-where}) into (\ref{marginal-GLMB}), i.e.,
\begin{equation}\label{Pro4-proof-1}
\begin{split}
  \bpi_{\bPsi_1}(\mathbf{X})&=  \sum_{I_2\in\mathcal{F}(\mathbb{L}/\mathbb{L}_1)}\sum_{I\in\mathcal{F}(\mathbb{L})}\omega^{(I)}\delta_{I}(\mathcal{L}(\bX)\cup I_2) {[p^{(I)}]}^\bX \\
&=\sum_{I_2\in\mathcal{F}(\mathbb{L}/\mathbb{L}_1)}\sum_{I_1\in\mathcal{F}(\mathbb{L}_1)}\delta_{I_1}(\mathcal{L}(\bX))\omega^{(I_1\cup I_2)}{[p^{(I)}]}^\bX.
\end{split}
\end{equation}
\end{proof}
\section{Proof of Proposition 6}
\begin{proof}
Write the LMB density of  form (\ref{LMB}) in GLMB form
\begin{equation}
\bpi_{\text{LMB}}(\bX)=\Delta(\bX)\sum_{I\in\mathcal{F}(\mathbb{L})}\omega(I)\delta_{I}(\mathcal{L}(\bX))p^\bX
\end{equation}
with
\begin{equation}\label{LMB-where}
\begin{split}
\mathbb{C}=&\mathcal{F}(\mathbb{L})\\
\omega^{(c)}(L)=&\omega(I)\delta_I(L)\\
p^{(c)}=&p.
\end{split},
\end{equation}
Hence, according to Proposition 3, the set marginal density of $\bPsi_1$ with respect to the LMB RFS $\bPsi$ can be obtained  by substitution of (\ref{LMB-where}) into (\ref{marginal-GLMB}), i.e.,
\begin{equation}\label{Pro4-proof-1}
\begin{split}
  \bpi_{\bPsi_1}(\mathbf{X})&=  \sum_{I_2\in\mathcal{F}(\mathbb{L}/\mathbb{L}_1)}\sum_{I\in\mathcal{F}(\mathbb{L})}\omega^{(I)}\delta_{I}(\mathcal{L}(\bX)\cup I_2) {p}^\bX \\
&=\sum_{I_2\in\mathcal{F}(\mathbb{L}/\mathbb{L}_1)}\sum_{I_1\in\mathcal{F}(\mathbb{L}_1)}\delta_{I_1}(\mathcal{L}(\bX))\omega^{(I_1\cup I_2)}{p}^\bX.
\end{split}
\end{equation}
\end{proof}

\bibliographystyle{plain}
\bibliography{refs}

\end{document}